\documentclass[aps,twocolumn,pra,twoside,amssymb,amsmath]{revtex4}
\usepackage{amssymb}
\usepackage{graphicx}
\usepackage{amsmath}
\usepackage{colordvi}
\usepackage{bbm}

\newcommand{\tr}{{\rm Tr}}

\begin{document}
\title{Not all entangled states are useful for ancilla-assisted quantum process tomography}
\author{Guo-Dong Lu}
\affiliation{School of Physical Science and Technology, Ningbo University, Ningbo, 315211, China}
\author{Zhou Zhang}
\affiliation{School of Physical Science and Technology, Soochow University, Suzhou, 215006, China}
\author{Yue Dai}
\affiliation{School of Physical Science and Technology, Soochow University, Suzhou, 215006, China}
\author{Yu-Li Dong}
\affiliation{School of Physical Science and Technology, Soochow University, Suzhou, 215006, China}
\author{Cheng-Jie Zhang}
\email{chengjie.zhang@gmail.com}
\affiliation{School of Physical Science and Technology, Ningbo University, Ningbo, 315211, China}
\affiliation{School of Physical Science and Technology, Soochow University, Suzhou, 215006, China}

\begin{abstract}
It is well known that one can extract all the information of an unknown quantum channel by means of quantum process tomography, such as standard quantum-process tomography and ancilla-assisted quantum process tomography (AAQPT). Furthermore, it has been shown that entanglement is not necessary for AAQPT, there exist separable states which are also useful for it. Surprisingly, in this work we find that not all entangled states are useful for AAQPT, there also exist some entangled states which are useless. The realignment operation  used in entanglement detection can be related to the question whether a bipartite state is useful for AAQPT. We derive the relationship between them and show the process of extracting the complete information of an unknown channel by the realignment operation. Based on this relationship, we present examples of a two-qutrit entangled state and a two-qutrit bound entangled state. Both of these two examples are entangled but they cannot be used for AAQPT. Last but not least, experimental verification has also been performed on the IBM platform.
\end{abstract}

\maketitle
\section{Introduction}
Entanglement is widely recognized as one of the most important resources in quantum information processing. Maximally entangled states have been applied to various information-processing tasks, including quantum communication channels \cite{CHB}, quantum cryptography \cite{AE}, quantum teleportation \cite{GBC}, and so on. Entanglement detection is one of the open questions in quantum information theory. Many entanglement detection criteria have been proposed, such as the partial transposition criterion \cite{pt,pt2}. The computable cross-norm or realignment (CCNR) criterion \cite{pp,oo} is a strong criterion which is powerful enough to detect some bound entangled states.

\begin{figure}
\begin{center}
\includegraphics[scale=0.45]{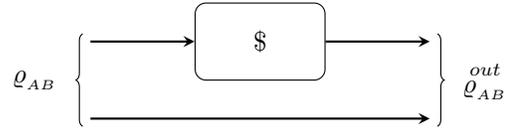}
\caption{The scheme of ancilla-assisted process tomography. In order to extract the complete information of quantum channel $\$$ in system $A$, an auxiliary system $B$ has been used to assist. If the input state $\varrho_{AB}$ is of faithfulness, the complete  information of channel $\$$ can then be obtained from the output state $\varrho_{AB}^{out}$.} \label{fig1}
\end{center}
\end{figure}

Quantum process tomography is a technique employed to fully characterize unknown quantum channels \cite{ov,og,of,oh,GP,os,DW,JB,ob,oc,od,oe,ou,oa,qpt1,qpt2,qpt3,qpt4,qpt5,qpt6,qpt7,qpt8,qpt9,qpt10,qpt11,qpt12,qpt13}.
Complete information of a quantum channel $\$$ in Eq. (\ref{Kraus}) is to determine a set of operation elements $\{K_n\}$ for $\$$, such that for an arbitrary input state $\sigma$ one can obtain the corresponding output state $\$(\sigma)$.
The currently known methods for completely obtaining channel information fall into three categories: standard quantum-process tomography (SQPT) \cite{og,of}, ancilla-assisted quantum process tomography (AAQPT) \cite{oh,GP,os,DW,JB} and direct characterization of quantum dynamics (DCQD) \cite{ob,oc,od,oe}. All SQPT methods are based on multiple comparisons between the output state of the channel and the appropriate input state to analyze the process impact of the identification channel operate. The AAQPT methods usually establish a corresponding relationship between the channel information and the quantum state with auxiliary systems. This correspondence allows us to derive the complete information of the channel, see Fig. \ref{fig1}. The methods of DCQD \cite{ou} are different from SQPT and AAQPT, which do not need to perform any state tomography.

The way of extracting channel information in Refs. \cite{GP,os} clearly belongs to AAQPT. They proposed a novel bipartite quantum states property named \textit{faithfulness} \cite{GP}, which indicated whether the complete information of any quantum channel can be obtained or not. The sufficient and necessary conditions of a bipartite state with faithfulness have been given in Refs. \cite{GP,os}. If the input state is of faithfulness, the corresponding output state carries complete information of the channel, and thus the input state is useful for AAQPT. In Ref. \cite{GP}, the authors showed that there exist separable states which can also be used for AAQPT. Interestingly, one may ask the related question, \textit{are all entangled states useful for AAQPT?}  We will focus on this question.

In this paper, we find that  not all entangled states are useful for AAQPT, there exist some entangled states which are useless. We first present that the realignment operator $\mathcal{R}$ in the CCNR criterion is related to the question whether a bipartite state is faithful or not. The relationship between the realignment operator and the faithfulness of a bipartite state is explained. Then we show the process of extracting the complete information of an unknown channel by the realignment operation. Based on this relationship, we present two examples of entangled states which are entangled but they cannot be used for AAQPT. Last but not least, experimental verification has also been performed on the IBM platform \textit{ibmq\_athens}.

\section{Entanglement is not sufficient for fully characterize quantum channels}
The output state $\$(\varrho)$, corresponding to an input state $\varrho$, of a linear, trace-preserving and completely positive map $\$$ \cite{oa} can be written in a so-called Kraus form \cite{om},
\begin{equation}\label{Kraus}
    \varrho \rightarrow \$(\varrho)=\sum_n K_n \varrho K_n^\dag,
\end{equation}
where $\{K_n\}$ are operators on the Hilbert space $\mathcal{H}$ of the quantum system and satisfy the completeness condition $\sum_n K_n^\dag K_n=I$.

The Choi-Jamio{\l}kowski isomorphism states that every superoperator $\$$ on $\mathcal{H}$ has a one-to-one correspondence to a state on $\mathcal{H}\otimes \mathcal{H}$ given by \cite{TJ,MD,or}
\begin{eqnarray}
S_\$:=\$\otimes I (|\Phi^+\rangle\langle\Phi^+|),\label{S}
\end{eqnarray}
where $|\Phi^+\rangle=\sum_i|ii\rangle$ is the unnormalized maximally entangled state. The inverse relation is
\begin{eqnarray}
\$(\varrho)=\tr_B[(I\otimes \varrho^T)S_\$],
\end{eqnarray}
where $\tr_B$ is the partial trace for the second $\mathcal{H}$, and $T$ is transposition.

From the Choi-Jamio{\l}kowski isomorphism, one can see that the state $S_\$$ in Eq. (\ref{S}) contains complete information of the quantum channel $\$$. One natural question is that: is maximally entangled state $|\Phi^+\rangle$ necessary? i.e., can we use another density matrix $\varrho_{_{AB}}$ on $\mathcal{H}\otimes \mathcal{H}$ such that $\$\otimes I (\varrho_{_{AB}})$ contains complete information of the quantum channel $\$$ as well? As shown in Fig. \ref{fig1} , Ref. \cite{GP} proved that $\$\otimes I (\varrho_{_{AB}})$ contains complete information of the quantum channel $\$$ if and only if all the singular values of $\check{R}(\varrho_{_{AB}})$ are nonzero (i.e. $\check{R}(\varrho_{_{AB}})^{-1}$ exists), where 
\begin{eqnarray}
\check{R}(\varrho_{_{AB}}):=(\varrho_{_{AB}}^{T_B}E)^{T_A},\label{checkR}
\end{eqnarray}
with  $E=\sum_{ij}|ij\rangle\langle ji|$ being the swap operator.

Interestingly, we find that the realignment operation $\mathcal{R}$ in the CCNR criterion is related to $\check{R}$. Thus, it has the similar consequence as $\check{R}$ in imprinting complete information of quantum channel on output states.

\textit{Theorem 1.}
The singular values of $\check{R}(\varrho_{_{AB}})$  are exactly the same as the ones of $\mathcal{R}(\varrho_{_{AB}})$. Thus, $\$\otimes I (\varrho_{_{AB}})$ contains complete information of the quantum channel $\$$, if and only if all the singular values of $\mathcal{R}(\varrho_{_{AB}})$ are nonzero (i.e. $\mathcal{R}(\varrho_{_{AB}}))^{-1}$ exists).

\textit{Proof.---}
Consider a general bipartite finite-dimensional state, $\varrho_{_{AB}}$ given by
\begin{eqnarray}
\varrho_{_{AB}}=\sum_{i,j,k,l}\varrho_{ij,kl}|i\rangle\langle j|\otimes|k\rangle\langle l|.
\end{eqnarray}
According to Eq. (\ref{checkR}), $\check{R}(\varrho_{_{AB}})$ can be expressed as
\begin{eqnarray}
\check{R}(\varrho_{_{AB}})=(\varrho_{_{AB}}^{T_B}E)^{T_A}=\sum_{i,j,k,l}\varrho_{ij,kl}|k\rangle\langle i|\otimes|l\rangle\langle j|.
\end{eqnarray}

For the realignment operation, we can obtain $\mathcal{R}(\varrho_{_{AB}})$ based on the definition of the realignment operation in Ref. \cite{pp} (here, we use a equivalent realignment method compared with the original one in Ref.  \cite{pp}),
\begin{eqnarray}
\mathcal{R}(\varrho_{_{AB}})=\sum_{i,j,k,l}\varrho_{ij,kl}|i\rangle\langle k|\otimes|j\rangle\langle l|.
\end{eqnarray}

We can find the following relationship between $\check{R}(\varrho_{_{AB}})$ and $\mathcal{R}(\varrho_{_{AB}})$,
\begin{eqnarray}
[\check{R}(\varrho_{_{AB}})]^{T_{AB}}=\mathcal{R}(\varrho_{_{AB}}),
\end{eqnarray}
where $T_{AB}$ stand for a transpose of the A and B part of the matrix $\check{R}(\varrho_{_{AB}})$. Any matrix $\check{R}(\varrho_{_{AB}})$ can be expressed as the form of singular decomposition
\begin{eqnarray}
\check{R}(\varrho_{_{AB}})=U\Sigma V^{\dagger},
\end{eqnarray}
which $U$ and $V$ are unitary matrices, $\Sigma$ is a diagonal matrix with singular values of $\check{R}(\varrho_{_{AB}})$. Thus, $\mathcal{R}(\varrho_{_{AB}})$ can be expressed as
\begin{eqnarray}
\mathcal{R}(\varrho_{_{AB}})&=&[\check{R}(\varrho_{_{AB}})]^{T_{AB}}=[U\Sigma V^{\dagger}]^{T_{AB}}\nonumber\\
                            &=&V^\ast\Sigma^T U^T=V^\ast\Sigma U^T.
\end{eqnarray}
Therefore, $\check{R}(\varrho_{_{AB}})$ and $\mathcal{R}(\varrho_{_{AB}})$ have exactly the same singular values.  $\$\otimes I (\varrho_{_{AB}})$ contains complete information of the quantum channel $\$$, if and only if all the singular values of $\mathcal{R}(\varrho_{_{AB}})$ are nonzero (i.e. $\mathcal{R}(\varrho_{_{AB}}))^{-1}$ exists). \hfill  $\square$

Based on Theorem 1, we can easily prove the following theorem by using the realignment operation.

\textit{Theorem 2.} Not all entangled states are useful for ancilla-assisted quantum process tomography. There exist entangled states which cannot be used for extracting complete information of quantum channels.

\textit{Proof.---}
To prove this theorem, we will show that there exist  entangled states $\varrho_e$ such that $\mathcal{R}(\varrho_e)$ have at least one singular value being zero.

For the first example, consider the following two-qutrit state $\sigma_E$,
\begin{eqnarray}
\sigma_E&=&p\frac{(|00\rangle+|11\rangle)(\langle00|+\langle11|)}{2} \nonumber \\
       &&+(1-p)\frac{(|00\rangle+|22\rangle)(\langle00|+\langle22|)}{2},
\end{eqnarray}
which is entangled when $0\leq p\leq1$. One can easily obtain
\begin{equation}
\mathcal{R}(\sigma_E)=\mathrm{diag}\bigg\{ \frac{1}{2},\frac{p}{2},\frac{1-p}{2},\frac{p}{2},\frac{p}{2},0,\frac{1-p}{2},0,\frac{1-p}{2}\bigg\},
\end{equation}
then $\mathcal{R}(\sigma_E)$ has at least two zero singular values. Thus, the entangled state $\sigma_E$ cannot be used for extracting complete information of quantum channels on qutrit states, since for all values of $p\in[0,1]$ the value  $\mathcal{R}(\sigma_E)^{-1}$ does not exist.

Another example of entangled state, we review a 3$\times$3 bound entangled state \cite{bound},
\begin{eqnarray}
\rho=\frac{1}{8a+1}
\left(
\begin{array}{ccccccccc}
 a & 0 & 0 & 0 & a & 0 & 0 & 0 & a\\
 0 & a & 0 & 0 & 0 & 0 & 0 & 0 & 0\\
 0 & 0 & a & 0 & 0 & 0 & 0 & 0 & 0\\
 0 & 0 & 0 & a & 0 & 0 & 0 & 0 & 0\\
 a & 0 & 0 & 0 & a & 0 & 0 & 0 & a\\
 0 & 0 & 0 & 0 & 0 & a & 0 & 0 & 0\\
 0 & 0 & 0 & 0 & 0 & 0 & b & 0 & c\\
 0 & 0 & 0 & 0 & 0 & 0 & 0 & a & 0\\
 a & 0 & 0 & 0 & a & 0 & c & 0 & b
\end{array}
\right),
\end{eqnarray}
where $0<a<1$, $b=(1+a)/2$, and $c=\sqrt{1-a^2}/2$.
After the realignment operation,
one can find that the $\mathcal{R}(\rho)$ has one singular value of zero, thus $\mathcal{R}(\rho)^{-1}$ does not exist. According to the previous derivation, this state $\rho$ cannot obtain the complete information of the channel although the state is an entangled state.

Therefore, not all entangled states are useful for AAQPT. There exist entangled states which cannot be used for extracting complete information of quantum channels.
\hfill  $\square$

In Ref. \cite{GP}, it has been shown that entanglement is not necessary for extracting complete information of quantum channels. Surprisingly, entanglement is not sufficient either as proved in Theorem 2. Thus, one can conclude that entanglement is neither necessary nor sufficient for extracting complete information of quantum channels.

\section{Extracting complete information of quantum channels from $\mathcal{R}(\varrho_{_{AB}})$}
In the above section, Theorem 1 only proved that one can obtain the complete information of quantum channels from $\mathcal{R}(\varrho_{_{AB}})$ if and only if $[\mathcal{R}(\varrho_{_{AB}})]^{-1}$ exists, but the method for extracting complete information of quantum channels from $\mathcal{R}(\varrho_{_{AB}})$ is not given.
In this section, we give a general method for extracting complete information of quantum channels from $\mathcal{R}(\varrho_{_{AB}})$.
The method is derived as follows.

For any input state $\sigma$, we have the corresponding output state $\$(\sigma)$,
\begin{eqnarray}
\$(\sigma)=\sum_n K_n \sigma K_n^\dag.\label{M16}
\end{eqnarray}
As introduced in Ref. \cite{GP}, we define $|\sigma\rangle$ as follows,
\begin{eqnarray}
|\sigma\rangle:=\sigma\otimes I \sum_i|ii\rangle= I\otimes \sigma^T\sum_i|ii\rangle.\label{rangle}
\end{eqnarray}
In the same way, we can also define $\langle\sigma|$ as follows,
\begin{eqnarray}
\langle\sigma|:=\sum_i\langle ii|\sigma\otimes I.
\end{eqnarray}
Now we can  express the realignment operator $\mathcal{R}$ as \cite{zhang,zhang2}
\begin{eqnarray}
\mathcal{R}(\sigma_1\otimes\sigma_2)=|\sigma_1\rangle\langle\sigma_2^*|.
\end{eqnarray}
Thus, according to Eq. (\ref{rangle}), $|\$(\sigma)\rangle$ can be expressed as
\begin{eqnarray}
|\$(\sigma)\rangle
&=&\sum_n K_n \sigma K_n^\dag\otimes I \sum_i|ii\rangle\nonumber\\
&=&\sum_n K_n \sigma\otimes (K_n^\dag)^T\sum_i|ii\rangle\nonumber\\
&=&(\sum_n K_n\otimes K_n^\ast) (\sigma\otimes I)\sum_i|ii\rangle\nonumber\\
&=&\sum_n K_n\otimes K_n^\ast|\sigma\rangle,\label{M}
\end{eqnarray}
where the second equation holds since $K_n^\dag\otimes I \sum_i|ii\rangle=I \otimes (K_n^\dag)^T\sum_i|ii\rangle$.

For an arbitrary finite-dimensional bipartite state $\varrho_{_{AB}}$, we can expand it under local orthonormal operator bases \cite{os,zhang,LOO},
\begin{eqnarray}
\varrho_{_{AB}}=\sum_k\lambda_k G_k^A\otimes G_k^B,\label{LOO}
\end{eqnarray}
which $\{G_k^A\}$ and $\{G_k^B\}$ satisfy the following orthogonal conditions,
\begin{eqnarray}
\tr(G_k^A G_{k^\prime}^A)=\tr(G_k^B G_{k^\prime}^B)=\delta_{k{k^\prime}}.
\end{eqnarray}
After the realignment operation, it can be expressed as
\begin{eqnarray}
\mathcal{R}(\varrho_{_{AB}})=\sum_k\lambda_k |G_k^A\rangle \langle (G_k^B)^*|.
\end{eqnarray}
Based on Eq. (\ref{LOO}), $\$\otimes I (\varrho_{_{AB}})$  can be represented as
\begin{eqnarray}
\$\otimes I (\varrho_{_{AB}})=\sum_k\lambda_k \$(G_k^A)\otimes G_k^B.
\end{eqnarray}
Therefore, $\mathcal{R}(\$\otimes I (\varrho_{_{AB}}))$ can be written as
\begin{eqnarray}
\mathcal{R}\big(\$\otimes I (\varrho_{_{AB}})\big)&=&
    \sum_k\lambda_k |\$(G_k^A)\rangle\langle (G_k^B)^*|\nonumber\\
    &=&\sum_k\lambda_k (\sum_n K_n\otimes K_n^\ast) |G_k^A\rangle\langle (G_k^B)^*|\nonumber\\
    &=&M\mathcal{R}(\varrho_{_{AB}}),
\end{eqnarray}
where we have defined
\begin{equation}\label{}
  M:=\sum_n K_n\otimes K_n^\ast,\label{Mdefine}
\end{equation}
and used \cite{GP}
\begin{equation}\label{}
 |\$(G_k^A)\rangle=|\sum_n K_n G_k^A K_n^\dag\rangle=\sum_n K_n\otimes K_n^\ast |G_k^A\rangle,
\end{equation}
and $M$ has complete information of the quantum channel.

\begin{figure*}[hbtp]
\begin{center}
\includegraphics[scale=1]{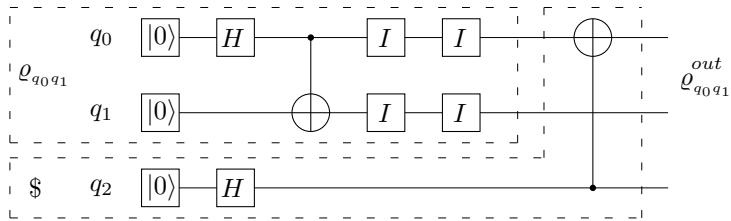}
\caption{ The initial state of $q_0$ , $q_1$ and $q_2$ is $|0\rangle$. $H$ is the Hadamard Gate ($H$-Gate) and $I$ stands for the identity gate, which has no theoretical influence on the input state $\varrho_{q_0q_1}$, but practically, it will have some interference to cause some errors in the input state. The quantum channel $\$$ has been realized by an
auxiliary qubit $q_2$. By the CNOT gate $U^{cnot}_{q_2q_0}$, we can realize the channel (\ref{Pauli}) after tracing the qubit $q_2$.} \label{fig2}
\end{center}
\end{figure*}

Therefore, if and only if all singular values of $\mathcal{R}(\varrho_{_{AB}})$ are nonzero \big(i.e. $\mathcal{R}(\varrho_{_{AB}})^{-1}$ exists\big), one can obtain $M$  which contains complete information of the channel, and it can be expressed as
\begin{eqnarray}
M=\mathcal{R}\big(\$\otimes I (\varrho_{_{AB}})\big)\mathcal{R}(\varrho_{_{AB}})^{-1}.\label{M27}
\end{eqnarray}
When the arbitrary input state is $\sigma$ and the corresponding output state is $\$(\sigma)$, they have the following relationship based on Eq. (\ref{M}),
\begin{eqnarray}
|\$(\sigma)\rangle=M|\sigma\rangle=\mathcal{R}\big(\$\otimes I (\varrho_{_{AB}})\big)\mathcal{R}\big(\varrho_{_{AB}}\big)^{-1}|\sigma\rangle.\label{M28}
\end{eqnarray}
By solving $M$, we can get the matrix information of $\$(\sigma)$ after transformation.

Because of $|\sigma\rangle=\sigma\otimes I \sum_i|ii\rangle$, we can get $\sigma$ reversely,
\begin{eqnarray}
\sigma=\tr_B\big(|\sigma\rangle\sum_i\langle ii|\big).
\end{eqnarray}
Similarly, for $|\$(\sigma)\rangle=M|\sigma\rangle$, we can also get $\$(\sigma)$,
\begin{eqnarray}
\$(\sigma)=\tr_B\big(|\$(\sigma)\rangle\sum_i\langle ii|\big)=\tr_B\big(M|\sigma\rangle\sum_i\langle ii|\big).\label{output}
\end{eqnarray}
Thus, for an arbitrary input state $\sigma$, we first get $M$ and $|\sigma\rangle$, then based on Eq. (\ref{output}) we can obtain the output state
$\$(\sigma)$.

\section{quantum circuits simulation}
In this section, we will design a quantum circuit, and compare the  experimental result of the channel information obtained by the realignment operator $\mathcal{R}$ with the theoretical result of this channel. We realize this quantum circuit on IBM quantum processor \textit{ibmq\_athens}. It is a five-qubit quantum system, we only use three qubits of them.

The quantum circuit is shown in Fig. \ref{fig2}. We denote $\varrho_{q_0q_1}$ and $\varrho _{q_0q_1}^{out}$ as input and output states, respectively. The input state  $\varrho_{q_0q_1}$  is a two-qubit state composed of $q_0$ and $q_1$. The initial states of $q_0$ and $q_1$ are $|0\rangle$. The quantum channel $\$$ has been realized by an
auxiliary qubit $q_2$, which can be expressed as follows,
\begin{eqnarray}
\$(\sigma)=\sum_{n=1}^2 K_n \sigma K_n^\dag,\label{Pauli}
\end{eqnarray}
where $K_1=I/\sqrt{2}$ and $K_2=\sigma_x/\sqrt{2}$. Based on Eq. (\ref{Mdefine}), one can get the theoretical value of $M$ which contains complete information of the channel $\$$,  
\begin{eqnarray}
M&=&\sum_{n=1}^2 K_n\otimes K_n^\ast
=\frac{I}{\sqrt{2}}\otimes \frac{I^\ast}{\sqrt{2}}+\frac{\sigma_x}{\sqrt{2}} \otimes \frac{\sigma^\ast_x}{\sqrt{2}}.
\label{thM}
\end{eqnarray}
From Fig. \ref{fig2}, one can calculate the theoretical results of density matrices for the input state $\varrho_{q_0q_1}$ and the output state $\varrho _{q_0q_1}^{out}$. It is worth noticing that $|\psi_{in}\rangle
=U^{cnot}_{q_0q_1}|+\rangle\otimes|0\rangle
=\frac{1}{\sqrt{2}}(|00\rangle+|11\rangle)$ and
\begin{eqnarray}
\varrho_{q_0q_1}=|\psi_{in}\rangle\langle \psi_{in}|.\label{rhoin}
\end{eqnarray}
For the input state after the $H$ Gate and the CNOT Gate, we have some $I$ operations. Here, $I$ Gates have no theoretical influence on the input state $\varrho_{q_0q_1}$. But for practical quantum computation, it will have some interference to cause some errors in the input state. As mentioned in the previous introduction of the quantum circuit, we reduce the purity of the input state to verify whether $\mathcal{R}$ operation still has a good effect of channel information extraction when the purity of the input state is not high.

From Fig. \ref{fig2}, the output state $\varrho _{q_0q_1}^{out}$ can be calculated as
\begin{eqnarray}
\varrho _{q_0q_1}^{out}
&=&\tr_{q_2}|\psi_{out}\rangle\langle\psi_{out}|\nonumber\\
&=&\frac{1}{2}|\psi_{in}\rangle\langle\psi_{in}|+\frac{1}{2}\sigma_x\otimes I|\psi_{in}\rangle\langle\psi_{in}|\sigma_x\otimes I,
\end{eqnarray}
where $|\psi_{out}\rangle
=U^{cnot}_{q_2q_0}|\psi_{in}\rangle\otimes|+\rangle
=1/\sqrt{2}|\psi_{in}\rangle\otimes |0\rangle+1/\sqrt{2}\sigma_x\otimes I|\psi_{in}\rangle\otimes |1\rangle$.

Our quantum circuit has been executed on IBM quantum processor \textit{ibmq\_athens} for $10,240$ times, which were divided into 10 batches. For each batch, the density matrices of the experimental input and output states have been obtained based on quantum state tomography, and the error bars of the following fidelities depicted  thrice of  the standard deviation of those 10 batches.
Thus, we obtained the practical channel information $M$ through Eq. (\ref{M27}) by using the practical density matrices of  $\varrho_{q_0q_1}$ and $\varrho _{q_0q_1}^{out}$.  In addition, we can calculate the fidelity of the input state between the theoretical value and the practical value. The fidelity is defined as follows \cite{of},
\begin{equation}
F(\varrho,\tilde{\varrho}):= \tr\sqrt{\varrho^{1/2}\tilde{\varrho}\varrho^{1/2}}.
\end{equation}
From Eq. (\ref{rhoin}), one can figure out that
\begin{equation}
F_{\varrho_{in}}=0.974\pm0.011.
\end{equation}
The fidelity of output state between the theoretical value and the practical value can also be obtained,
\begin{equation}
F_{\varrho_{out}}=0.954\pm0.027.
\end{equation}

In order to compare the  theoretical value of $M$ in Eq. (\ref{thM}) and the practical value of $M$  through Eq. (\ref{M27}), we use $|0\rangle$, $|1\rangle$, $|+\rangle$, $|-\rangle$, $|L\rangle$ and $|R\rangle$ as input states, where $|+/-\rangle=(|0\rangle\pm|1\rangle)/\sqrt{2}$ and $|L/R\rangle=(|0\rangle\pm i|1\rangle)/\sqrt{2}$. We calculate the fidelities between the output states from the  theoretical value $M$ and the  practical value $M$,
\begin{eqnarray}
F_{out_{|0\rangle}}&=&0.999\pm0.004;\
F_{out_{|1\rangle}}=0.997\pm0.007;\nonumber\\
F_{out_{|+\rangle}}&=&0.968\pm0.050;\
F_{out_{|-\rangle}}=0.974\pm0.050;\nonumber\\
F_{out_{|L\rangle}}&=&0.998\pm0.002;\
F_{out_{|R\rangle}}=0.998\pm0.005.\nonumber
\end{eqnarray}

Although the actual measured value deviates from the theoretical value due to systematic or measurement errors when $\mathcal{R}$ is used to obtain information for the channel, its fidelity is still good to a certain extent. And the error is understandable.

\section{Discussions and conclusions}
We discuss some other possible roles of the operator $\check{R}$ in the Ref. \cite{GP}. It is well known that the realignment operation $\mathcal{R}$ is often used as the criterion of entanglement of quantum states. If a state is separable then the sum of all singular values of $\mathcal{R}(\varrho_{AB})$ is less than 1, and it is entangled when the sum of the singular values is greater than 1 \cite{pp,oo,op}. From Theorem 1, since the singular values of quantum states are same when they are acted on by $\check{R}$ and $\mathcal{R}$, we conclude that $\check{R}$ can also be used  as a criterion for entanglement of quantum states.

Moreover, we have presented an example of 3$\times$3 bound entangled state, which is not faithful. Is it possible to prove that all the bound entangled states are unfaithful? If it is true, we can see that the set of bound entangled states is the subset of all unfaithful states. This is an interesting open question for our further research.

In Ref. \cite{os}, the authors present another necessary and sufficient condition for faithful states: A state $\sigma$ of $AB$ may be used to perform AAQPT if and
only if the Schmidt number of $\sigma$ is $d_A^2$, where $d_A$ is the dimension of the state space of system $A$. We can see that $\mathcal{R}(\sigma)$ is equivalent to the Schmidt
decomposition of $\sigma$ and the Schmidt number of $\sigma$ is equal to the number of the number of nonzero singular values of $\mathcal{R}(\sigma)$. Therefore, our condition ($\mathcal{R}(\sigma)^{-1}$ exists) is equivalent to the condition shown in Ref. \cite{os} as well.

In conclusion, it is found in Ref. \cite{GP} that the input state $\varrho_{AB}$ is faithful if $\check{R}(\varrho_{AB})$ is invertible, i.e., none of the singular values of $\check{R}(\varrho_{AB})$ are zero. In this paper, we find that $\mathcal{R}(\varrho_{AB})$ and $\check{R}(\varrho_{AB})$ have equal consequence on the solution of the singular value. By using $\mathcal{R}(\varrho_{AB})$, we can also get the complete channel information just like $\check{R}(\varrho_{AB})$. Furthermore, through this property, we use two entangled state to prove that not all entangled states are useful for AAQPT.

\section*{ACKNOWLEDGMENTS}
This work is supported by the National Natural Science Foundation of China (Grant No. 11734015), and K.C. Wong Magna Fund in Ningbo University.

\end{document}